\documentclass[prd,aps,preprint,nofootinbib]{revtex4}
\usepackage{epsfig}
\usepackage{color}
\usepackage{axodraw}

\usepackage{amsmath}

\begin{document}


\title{Single Spin Asymmetries in Heavy Quark and Antiquark
Productions}

\author{Feng Yuan}
\email{fyuan@lbl.gov} \affiliation{Nuclear Science Division,
Lawrence Berkeley National Laboratory, Berkeley, CA
94720}\affiliation{RIKEN BNL Research Center, Building 510A,
Brookhaven National Laboratory, Upton, NY 11973}
\author{Jian Zhou}
\email{jzhou@lbl.gov}
\affiliation{
Nuclear Science Division,
Lawrence Berkeley National Laboratory, Berkeley, CA
94720}\affiliation{Department of Physics, Shandong University, Jinan, Shandong 250100, China}

\begin{abstract}
The single transverse spin asymmetries in heavy quark and anti-quark
production from the quark-antiquark annihilation
channel contribution is studied by taking into account the initial
and final state interactions effects. Because of the different
color charges, the final state interaction effects lead to about
a factor of 3 difference in the spin asymmetry for heavy quark
over that for the anti-quark in the valence region of
low energy $pp$ collisions.
The experimental study of this model-independent prediction
shall provide a crucial test for the underlying mechanism for the
single spin asymmetry phenomena.

\end{abstract}

\maketitle

\newcommand{\be}{\begin{equation}}
\newcommand{\ee}{\end{equation}}
\newcommand{\ben}{\[}
\newcommand{\een}{\]}
\newcommand{\beqn}{\begin{eqnarray}}
\newcommand{\eeqn}{\end{eqnarray}}
\newcommand{\Tr}{{\rm Tr} }

{\bf 1. Introduction.} Single-transverse spin asymmetry (SSA) is
a novel and long standing phenomena in hadronic reactions~\cite{Anselmino:1994gn,{Qiu:1991pp}},
and has attracted intensive interests from both experiment and theory sides in the
last few years. It is a transverse spin dependent observable,
defined as the ratio of the cross section difference when we flip the
transverse spin of one of the hadrons involved in the scattering
over the sum of the cross sections,
$A_N\propto (d\sigma(S_\perp)-d\sigma(-S_\perp))/(d\sigma(S_\perp)+d\sigma(-S_\perp))$.
The experimental observation of SSAs in semi-inclusive
hadron production in deep inelastic scattering (SIDIS), in inclusive hadron
production in $pp$ scattering at collider energy at RHIC, and the
relevant azimuthal asymmetric distribution of hadron production in $e^+e^-$ annihilation
have motivated the theoretical developments in the last few years.
It was also argued that the SSA phenomena is closely related
to the orbital motion of quarks and gluons in the nucleon~\cite{Brodsky:2002cx,Ji:2002xn,{Burkardt:2004ur}}.

Theoretically, it has been realized that the initial/final state
interactions are crucial to leading to a nonzero SSA
in high energy scattering~\cite{Qiu:1991pp,Brodsky:2002cx}.
These effects also invalidate the naive-time-reversal invariance
argument~\cite{Collins:1992kk,Collins:2002kn,Ji:2002aa,{Boer:2003cm}}
for the non-existence of the so-called
Sivers function~\cite{Sivers:1990fh}, which describes a correlation
between the intrinsic transverse momentum of parton and the
transverse polarization vector of the nucleon.
The initial/final state interaction effects introduce
a process-dependence in the SSA observables.
For example, the SSA in the SIDIS process comes from final state
interaction whereas that in the Drell-Yan lepton pair production
comes from the initial state interaction. The consequence of this
difference is that there is a sign change between the SSAs for
these two processes~\cite{Collins:2002kn,{Ji:2006ub}}.
It is of crucial to test this nontrivial QCD
predictions by comparing the SSAs in these two processes.
The Sivers single spin asymmetry in SIDIS process has been
observed by the HERMES collaboration~\cite{hermes},
and the planned Drell-Yan measurement at RHIC and other
facility will test this prediction~\cite{dy-rhic}.

In this paper, we study another interesting probe for
the initial/final state interaction effects: the SSAs in
heavy quark and antiquark production in hadronic
process. Because the heavy quark and antiquark can be detected by their
decay products, their SSAs can be measured separately.
The heavy quark and antiquark produced in short distance partonic
processes will experience different final state interactions with the nucleon
remanet due to their different color charges, and therefore
the SSAs for heavy quark and antiquark will be different.
Detailed calculations in the following show that the difference could be as
large as a factor of 3 if the quark-antiquark channel contribution dominates.
This is a direct consequence of the final state interaction effects
in these SSAs.

The rest of the paper is organized as follows. In Sec.{\bf 2}, we will set
up our framework, and calculate the SSAs for heavy
quark and antiquark in the twist-three quark-gluon correlation
approach~\cite{Qiu:1991pp,new} and the initial/final state interactions effects are properly
taken into account. We further present some numerical results for experiments
at low energy $pp$ scattering where the quark-antiquark channel dominates,
and we will observe a factor of 3 difference between heavy quark and antiquark SSAs.
We conclude our paper in Sec. {\bf 3}.

{\bf 2. SSAs in Heavy Quark and Antiquark Productions.}
In order to obtain a non-zero SSA in high energy process,
we must generate a phase
from parton re-scattering in hard process~\cite{Qiu:1991pp,{Brodsky:2002cx}}.
Similar to other hadronic processes, in the heavy quark production in $pp$ collisions~\cite{Qiu:1991pp,new},
the partons involved in the re-scattering can come from the initial state or final
state, and they are referred as the initial and final state interactions,
respectively. Because of the different color charges
they carried in the scattering process, the heavy quark and
antiquark will have different effects from the final state interactions.
The initial state interactions, on the other hand, are the
same for both particles. In this paper,
we will take the quark-antiquark annihilation channel
as an example to demonstrate the unique consequence
from the different final state interaction effects on the
heavy quark and antiquark productions. The light quark
production in this channel has been studied
in~\cite{Qiu:1991pp,new}. We will extend to the heavy
quark production and study the relevant phenomenological consequence.

\begin{figure}[t]
\begin{center}
\includegraphics[width=13cm]{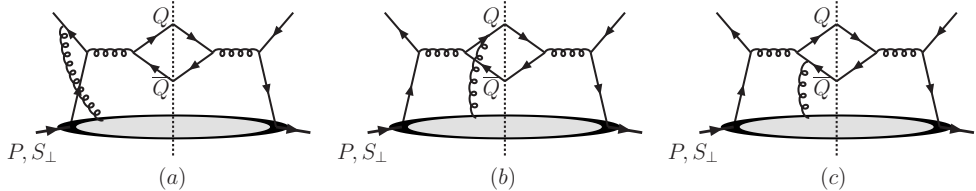}
\end{center}
\vskip -0.4cm \caption{\it Initial/final state interactions contributions
to the SSAs in heavy quark and antiquark productions: (a)
initial state interaction with color strength $1/2N_c$; (b) final state
interaction on the quark line with color strength $(N_c^2-2)/2N_c$;
(c) final state interaction on the antiquark line with color strength
$2/2N_c$. The mirror diagrams where the gluon
attaches to the right of the cut line are not shown, but also contribute.}
\end{figure}

In the quark-antiquark annihilation channel, the quark (or antiquark) from the polarized
nucleon annihilates the antiquark (or quark) from the unpolarized nucleon,
and produces a heavy quark and antiquark pair. For the spin-average
cross section, this forms the  $s-$channel partonic diagram.
As we mentioned above, the single transverse spin
dependent cross section comes from both initial and final
state interaction contributions. In Fig.~1 we show the relevant
initial (a) and final state interactions (b,c) diagrams.
Different from those in the SIDIS and Drell-Yan lepton pair
production, these initial and final state interactions
have different strength
because they have different color charges. This will result into
different color-factors for them, which can be carried out in
a model-independent way. For example, for the initial state
interaction of Fig.~1(a), we can label the incoming quark legs
from the polarized nucleon with color-index $i$ and $j$ ($(i,j)=1,3$
in the elementary representation in the $SU(3)$ gauge
group), the additional gluon with index $a$ ($a=1,8$
for the adjoint representation) connecting the nucleon remanet
with the partons in the hard partonic processes. The color-matrix
representing the coupling of
the nucleon with the partonic part will be uniquely the $SU(3)$ color matriices
$T^a_{ij}$, because of the gauge invariance. Therefore, the color-factor
associated with the initial state interaction can be formulated
as
\begin{equation}
{\rm Fig.1(a)} \propto (-){\rm tr}\left[T^aT^bT^aT^c\right]\times {\rm tr}\left[T^bT^c\right]=\frac{1}{2N_C}\frac{N_c^2-1}{4}\ ,
\end{equation}
where the additional minus sign is due to the initial state interaction effect.
Similarly, the final state interaction diagram of Fig.~1(b) will be
\begin{equation}
{\rm Fig.1(b)} \propto {\rm tr}\left[T^aT^bT^c\right]\times {\rm tr}\left[T^bT^aT^c\right]=\frac{N_C^2-2}{2N_C}\frac{N_c^2-1}{4}\ .
\end{equation}
Because this final state interaction is on the quark line, there is no additional minus sign. However,
the final state interaction on the antiquark line will have a minus sign,
\begin{equation}
{\rm Fig.1(c)} \propto (-){\rm tr}\left[T^aT^bT^c\right]\times {\rm tr}\left[T^aT^bT^c\right]=\frac{2}{2N_C}\frac{N_c^2-1}{4}\ .
\end{equation}
From the above results, we find that all the initial and final state
interactions contribute the same sign.
However, the strengths of these interactions are different. Especially, the two
final state interactions differ by  a factor of $3$.

Furthermore, the initial state interaction diagram Fig.~1(a) contributes
to the SSAs in both heavy quark and antiquark productions.
However, the final state interaction on the quark line Fig.~1(b) only contributes to the
heavy quark SSA, whereas that on the antiquark line Fig.~1(c)
only contributes to that for the antiquark. For example, when we study the SSA for
heavy quark production, the kinematics of heavy antiquark is
integrated out. The phase contribution from the final
state interaction on the antiquark line (Fig.~1(c)) will cancel out that from
the same diagram but with the gluon attachment to the right of
the cut line (the mirror diagram of Fig.~1(c))~\cite{Qiu:1991pp}. Therefore, the heavy quark
SSA has contribution from the initial state interaction of Fig.~1(a)
and the final state interaction on the quark line Fig.~1(b), whereas the heavy
antiquark SSA from the initial state interaction of Fig.~1(a) and
the final state interaction on the antiquark line Fig.~1(c).
If the final state interactions dominate
their SSAs, we will expect a factor of 3 difference
between the SSAs in heavy quark and antiquark
productions, because of the different strengths they
have as we have shown in the above analysis.
This is a model independent observation. Its test
shall provide a strong support for the underlying
physics for the SSA phenomena.

To calculate the SSA contributions from these diagrams,
it is appropriate  to adopt the twist-three quark-gluon-antiquark
correlation approach developed in~\cite{Qiu:1991pp}, because
the heavy quark mass provides a natural hard scale in the
computation of the perturbative diagrams, and the collinear
factorization approach is applicable. An important feature from
this approach is that the single spin asymmetry is naturally suppressed
in the limit of $P_\perp\ll M_Q$, where $P_\perp$ and  $M_Q$ are
transverse momentum and mass of the heavy quark.

The calculations are similar to those in~\cite{Qiu:1991pp,new}, where the
inclusive $\pi$ production in $p^\uparrow p\to \pi X$ was formulated.
The only difference is that we have to restore the mass dependence
for the final state quark and antiquark. The spin-average cross section
can be summarized as,
\begin{eqnarray}
E_Q\frac{d^3\sigma}{d^3P_Q}&=&\frac{\alpha_s^2}{S}\int
\frac{dx'}{x'}f_b(x')\frac{f_a(x)}{x}\frac{1}{x'S+(T-M_Q^2)}H^U_{ab\to Q\bar Q}(\tilde s,\tilde t,\tilde u) \ ,
\end{eqnarray}
where $P_Q$ is the heavy quark momentum, $S$ the hadronic center of mass energy,
$S=(P_A+P_B)^2$ with $P_A$ and $P_B$ the momenta for the incident polarized and
unpolarized nucleons,
$x$ and $x'$ are the momentum fractions carried by the incident partons, and $f_a$ and
$f_b$ are the associated parton distributions. The kinematic variables are defined
as: $E_Q=M_T\cosh y$ where $M_T=\sqrt{M_Q^2+P_\perp^2}$ and $y$ is the rapidity for
the heavy quark in the center of mass frame; $T=M_Q^2-M_T\sqrt{S}e^{-y}$ and
$U=M_Q^2-M_T\sqrt{S}e^{y}$; the partonic variable $\tilde s=xx' S$, $\tilde t=-xM_T\sqrt{S}e^{-y}$, and
$\tilde u=-x'M_T\sqrt{S}e^y$.
The leading contribution to heavy quark production comes from quark-antiquark
and gluon-gluon channels, and their hard factor $H^U$ are defined as,
\begin{eqnarray}
H^U_{q\bar q\to Q\bar Q}&=&\frac{N_c^2-1}{4N_c^2}\frac{2(\tilde t^2+\tilde u^2+2M_Q^2\tilde s)}{\tilde s^2}\ ,\\
H^U_{gg\to Q\bar Q}&=&\frac{2}{4N_c}\left(\frac{1}{\tilde t\tilde u}-\frac{2N_c^2}{N_c^2-1}\frac{1}{\tilde s^2}
\right)\frac{\tilde t\tilde u(\tilde t^2+\tilde u^2+4\tilde M_Q^2)-4\tilde s^2M_Q^4}{\tilde t\tilde u}\ .
\end{eqnarray}
In the twist-three framework~\cite{Qiu:1991pp}, the single transverse spin dependent
cross section  depends on the
twist-three quark-gluon correlation function, the so-called Qiu-Sterman
matrix element. They are defined as
\begin{eqnarray} \label{TF}
T_F^q(x_1,x_2) &\equiv &
\int\frac{d\zeta^-d\eta^-}{4\pi}e^{i(k_{q1}^+\eta^-+k_g^+\zeta^-)}
\, \epsilon_\perp^{\beta\alpha}S_{\perp\beta}
\label{eq1} \\
&&\left\langle PS|\overline\psi_q(0){\cal L}(0,\zeta^-)\gamma^+
g{F^+}_\alpha (\zeta^-){\cal L}(\zeta^-,\eta^-)
\psi_q(\eta^-)|PS\right\rangle \nonumber \ ,
\end{eqnarray}
where the sums over color and spin indices are implicit, $\epsilon^{\beta\alpha}\equiv \epsilon^{\mu\nu\beta\alpha}P_{A\mu}P_{B\nu}/P_A\cdot P_B$ with $\epsilon^{0123}=1$,
$|PS\rangle$ denotes the proton state, $\psi$ the quark field,
${F^+}_{\alpha}$ the gluon field tensor, and ${\cal L}$ represents the gauge link
to make the above definition gauge invariant. By summing the initial and
final state interaction contributions, the single spin dependent
differential cross section for heavy quark production can be written as
\begin{eqnarray}
E_Q\frac{d^3\Delta\sigma(S_\perp)}{d^3P_Q}&=&\frac{\alpha_s^2}{S}\int
\frac{dx'}{x'}f_{\bar q}(x')\frac{\epsilon^{\alpha\beta}S_\perp^\alpha
P_\perp^\beta}{x'S+(T-M_Q^2)}
\frac{1}{\tilde u}\left\{\left(\frac{T_F^q(x)}{x}-\frac{\partial}{\partial x}T_F^q(x)\right)\right.\nonumber\\
&&\times\left.
H_{q\bar q\to Q}+\frac{T_F^q(x)}{x}\tilde H_{q\bar q\to Q}\right\} \ ,
\end{eqnarray}
where $T_F^q(x)\equiv T_F^q(x,x)$. A similar expression can be obtained for the antiquark
by replacing $Q\to \bar Q$ in the above. The above hard factors include both initial and final
state interaction contributions.
The first term $H_{q\bar q\to Q}$ follows a compact formula containing
the derivative and non-derivative (of the twist-three function)
contributions~\cite{new}, whereas the second term $\tilde H_{q\bar q\to Q}$
only depends on the non-derivative of the twist-three correlation function and vanishes
in the massless limit $M_Q\to 0$. Under this limit, our results reproduce the
relevant formula in~\cite{new}. The hard factors are found,
\begin{eqnarray}
H_{q\bar q\to Q}&=&H^I_{q\bar q\to Q}+H^F_{q\bar q\to Q}(1+\frac{\tilde u}{\tilde t})\ ,\\
\tilde H_{q\bar q\to Q}&=&\tilde H^I_{q\bar q\to Q}+\tilde H^F_{q\bar q\to Q}(1+\frac{\tilde u}{\tilde t})\ ,
\end{eqnarray}
where
\begin{eqnarray}
H^I_{q\bar q\to Q}&=&H^I_{q\bar q\to \bar Q}=\frac{1}{4N_c^2}\frac{2(\tilde t^2+\tilde u^2+2M_Q^2\tilde s)}{\tilde s^2}, ~~\tilde H^I_{q\bar q\to Q}=\tilde H^I_{q\bar q\to \bar Q}=\frac{1}{N_c^2}\frac{M_Q^2}{\tilde s}\ ,\\
H^F_{q\bar q\to Q}&=&\frac{N_c^2-2}{4N_c^2}\frac{2(\tilde t^2+\tilde u^2+2M_Q^2\tilde s)}{\tilde s^2}, ~~\tilde H^F_{q\bar q\to Q}=\frac{N_c^2-2}{N_c^2}\frac{M_Q^2}{\tilde s}\ ,\\
H^F_{q\bar q\to \bar Q}&=&\frac{2}{4N_c^2}\frac{2(\tilde t^2+\tilde u^2+2M_Q^2\tilde s)}{\tilde s^2}, ~~\tilde H^F_{q\bar q\to \bar Q}=\frac{2}{N_c^2}\frac{M_Q^2}{\tilde s}\ ,
\end{eqnarray}
where the color factors have been discussed in the above.
The contributions from the antiquark-gluon correlation functions can
be obtained accordingly, by using the charge conjugation transformation.

It is important to note that the spin-average and spin-dependent cross section
contributions are both well defined in the limit of $P_\perp\to 0$, because
of the heavy quark mass. Therefore, the spin asymmetry defined as the ratio
of these two cross section terms vanishes when $P_\perp=0$. This is very different from
the massless particle production where the single spin asymmetry divergent
as $1/P_\perp$ as $P_\perp\to 0$~\cite{Qiu:1991pp,new}. Thus we can integrate out
the cross sections to small
transverse momentum region to obtain the spin asymmetry. In the following,
we will present the numerical results for the conventional
left-right asymmetry for the heavy quark production at low
energy $pp$ scattering. This asymmetry is defined as
\begin{equation}
A_N=\frac{L-R}{L+R}\ ,
\end{equation}
which counts the number difference between the left hand side and right hand side
for heavy quark production from the moving direction of the polarized (upward) nucleon.
It can be calculated from the cross section terms from above,
\begin{equation}
A_N=\frac{-\int_0^\pi d\phi d\Delta\sigma}{\int_0^\pi d\phi d\sigma} \ ,
\end{equation}
where $\phi$ is the azimuthal angle of heavy quark transverse momentum $P_\perp$
relative to the incoming nucleon transverse polarization vector $S_\perp$.
In the following, we will show the above asymmetry for heavy quark and antiquark
at low energy $pp$ scattering where the quark-antiquark channel dominates.
For the purpose of the demonstration, we adopt the paramterizations for the
twist-three Qiu-Sterman matrix elements in the valence region as~\cite{new},
\begin{equation}
T_F^a(x)=N_a x^{\alpha_a}(1-x)^{\beta_a} q_a(x) \ ,
\end{equation}
where $q_a(x)$ is the unpolarized quark distribution for flavor
$a$. The above parameters for the valence $u$ and $d$ quarks
were determined from a comparison to the single spin asymmetry data
in $p^\uparrow p\to \pi X$~\cite{new}, and they are also compatible with
the SIDIS data from HERMES~\cite{Yuan:2007zz}. For the unpolarized
parton distributions, we use the leading order (LO) CTEQ5L parton distribution
functions~\cite{cteq5l}.

\begin{figure}[t]
\begin{center}
\includegraphics[width=6cm]{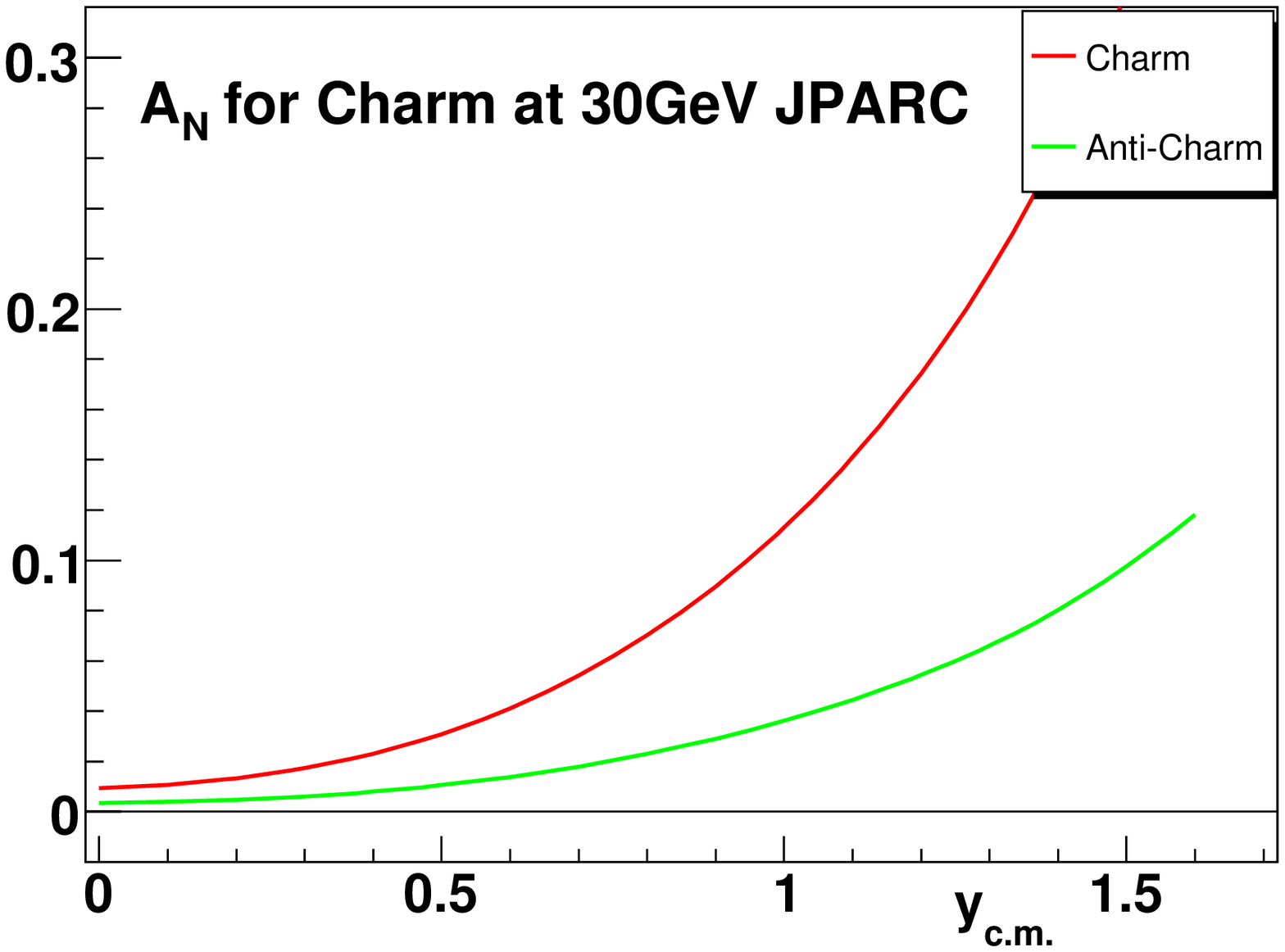}
\includegraphics[width=6cm]{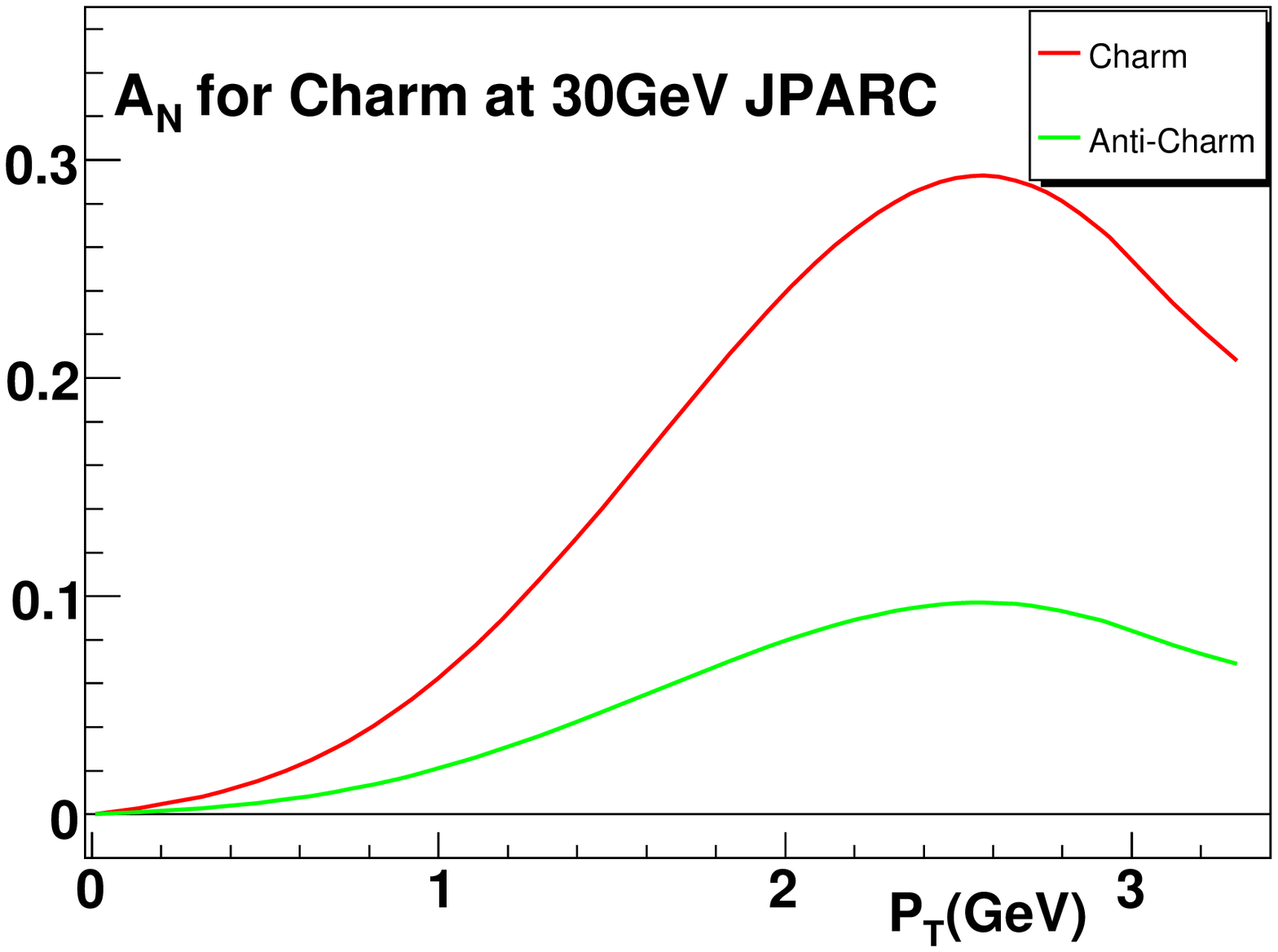}
\end{center}
\vskip -0.4cm \caption{\it Single spin asymmetries for charm
quark and anticharm quark production in $p^\uparrow p$ scattering
at JPARC energy: as functions of rapidities with $P_\perp$
integrated out (left panel);
and as functions of transverse momentum $P_\perp$ with
$y_{c.m.}>0$ (right panel).}
\end{figure}

In Fig.~2, we show our predictions for the heavy quark and antiquark
single spin asymmetries in $p^\uparrow p$ scattering
at JPARC energy region ($\sqrt{s}\approx 7.8 GeV$), where we expect
the quark-antiquark annihilation channel dominates the cross section.
The left panel shows the asymmetries $A_N$ as functions
of the center of mass rapidities of the charm quark and anti-charm quark
with the transverse momentum $P_\perp$ integrated out; the
right panel shows the asymmetries as functions of
the transverse momentum $P_\perp$ in the forward rapidity region ($y_{c.m.}>0$).
From these plots, we find that the asymmetries increase with
rapidity, similar to the general trend in other SSA observations~\cite{new}.
More importantly, the single spin asymmetry for the charm quark is much
larger than that for anticharm quark in the full rapidity range.
Similar observation was found for the asymmetries as functions
of the transverse momentum $P_\perp$. Both
asymmetries vanish as $P_\perp\to 0$ as we expected.

\begin{figure}[t]
\begin{center}
\includegraphics[width=6cm]{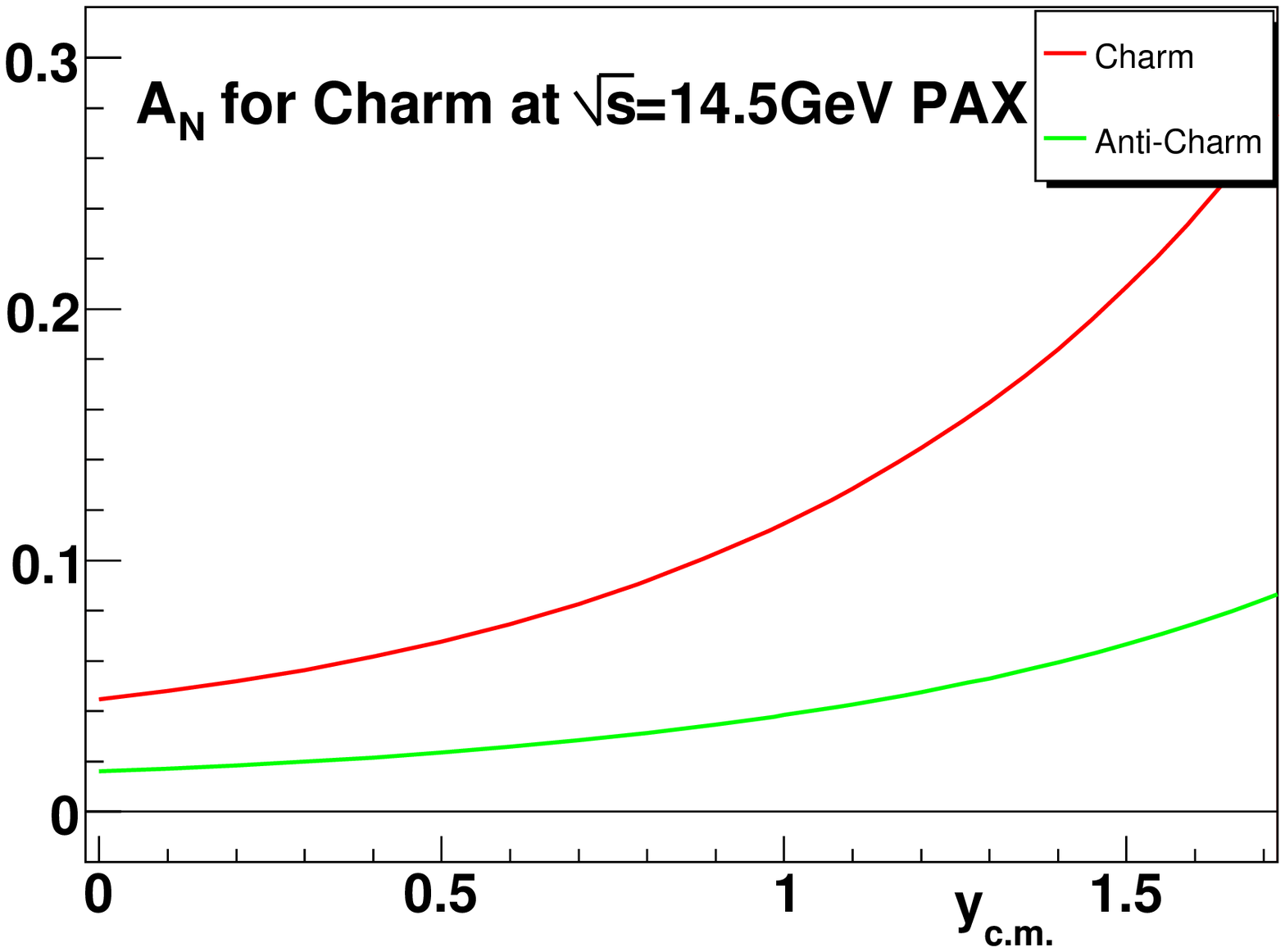}
\includegraphics[width=6cm]{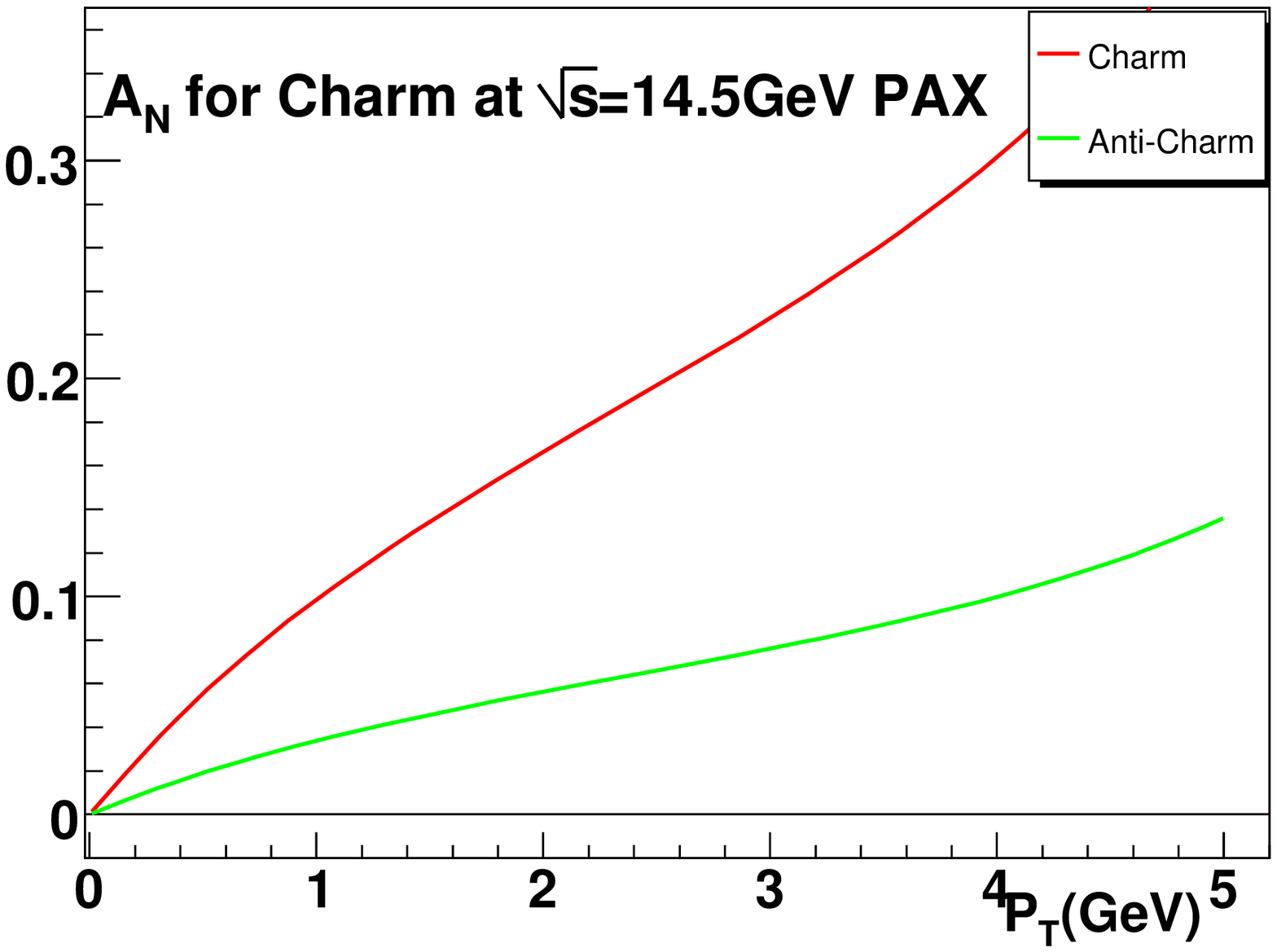}
\end{center}
\vskip -0.4cm \caption{\it Single spin asymmetries for charm and
anticharm quark production in $p^\uparrow \bar p$ scattering
at $\sqrt{s}=14GeV$ at PAX energy region: as functions
of raptidities with $P_\perp$ integrated out (left panel); as functions of charm quark
transverse momentum $P_\perp$ with $y_{c.m.}>0$ (right panel).}
\end{figure}

In Fig.~3, we show the predictions at $\sqrt{s}=14GeV$ $p^\uparrow \bar p$
scattering at the PAX experiment at GSI PAIR. Because
the quark-antiquark channel always dominates the cross section, we find
a sizable SSA for charm quark even at cental rapidity region. Again, we observe
that the SSA for charm quark is about a factor of $3$ larger than
that for the anticharm quark. In this experiment, we
can further study the spin asymmetry with the antiproton transversely
polarized $\bar p^\uparrow p\to Q(\bar Q)X$, where we expect the
SSA for anticharm quark is larger than that for the charm quark in the
forward direction of the polarized antiproton.
By using the charge conjugation transformation invariance, we will
have the following relation between the SSAs in these two
processes,
\begin{equation}
A_N(p^\uparrow \bar p\to C)=A_N(\bar p^\uparrow p\to \bar C)\ ,~~~
A_N(p^\uparrow \bar p\to \bar C)=A_N(\bar p^\uparrow p\to  C)\ ,
\end{equation}
for the same kinematics. From these relations, we can obtain the
corresponding results for $\bar p^\uparrow p$ experiments from
the plots in Fig.~3.

\begin{figure}[t]
\begin{center}
\includegraphics[width=6cm]{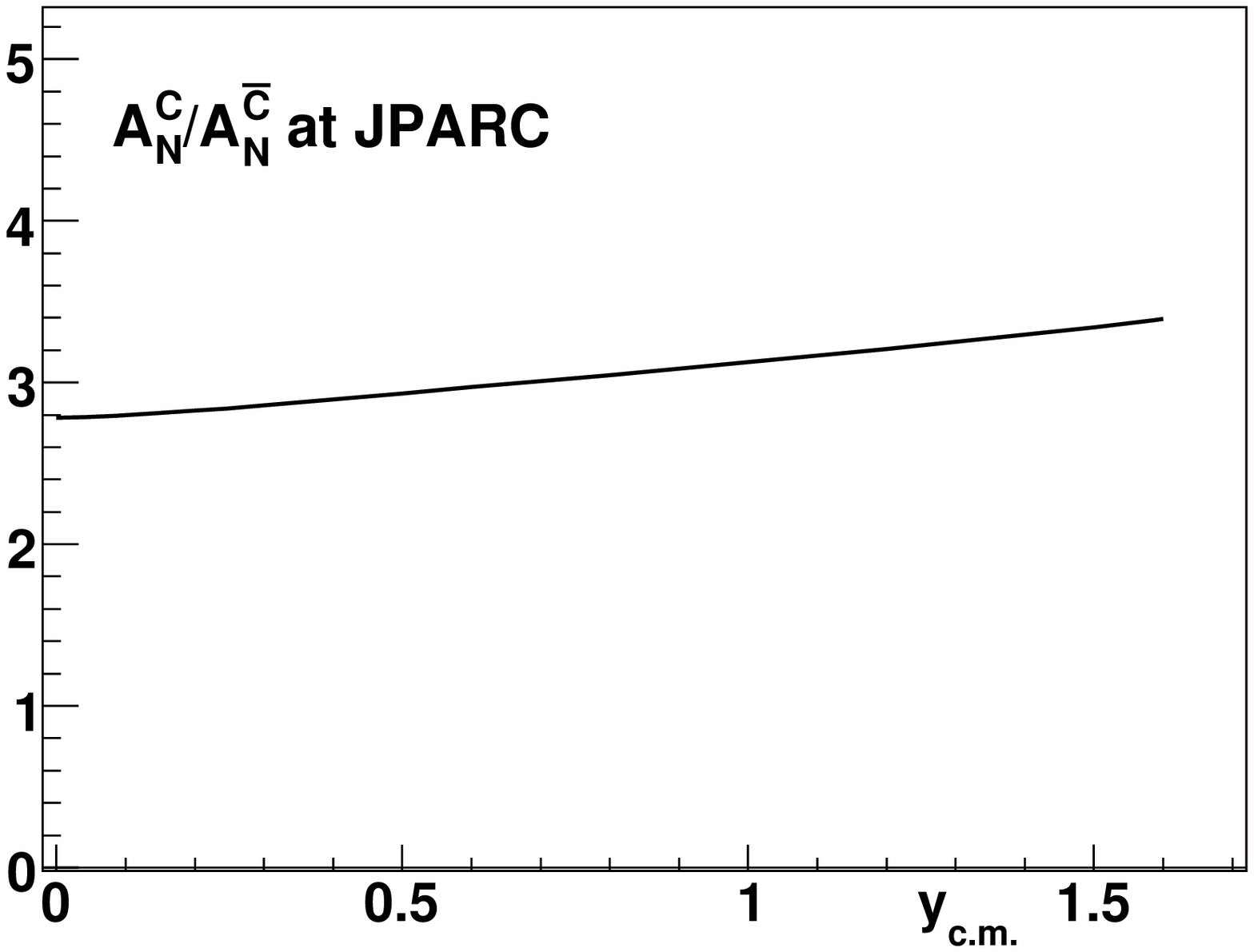}
\includegraphics[width=6cm]{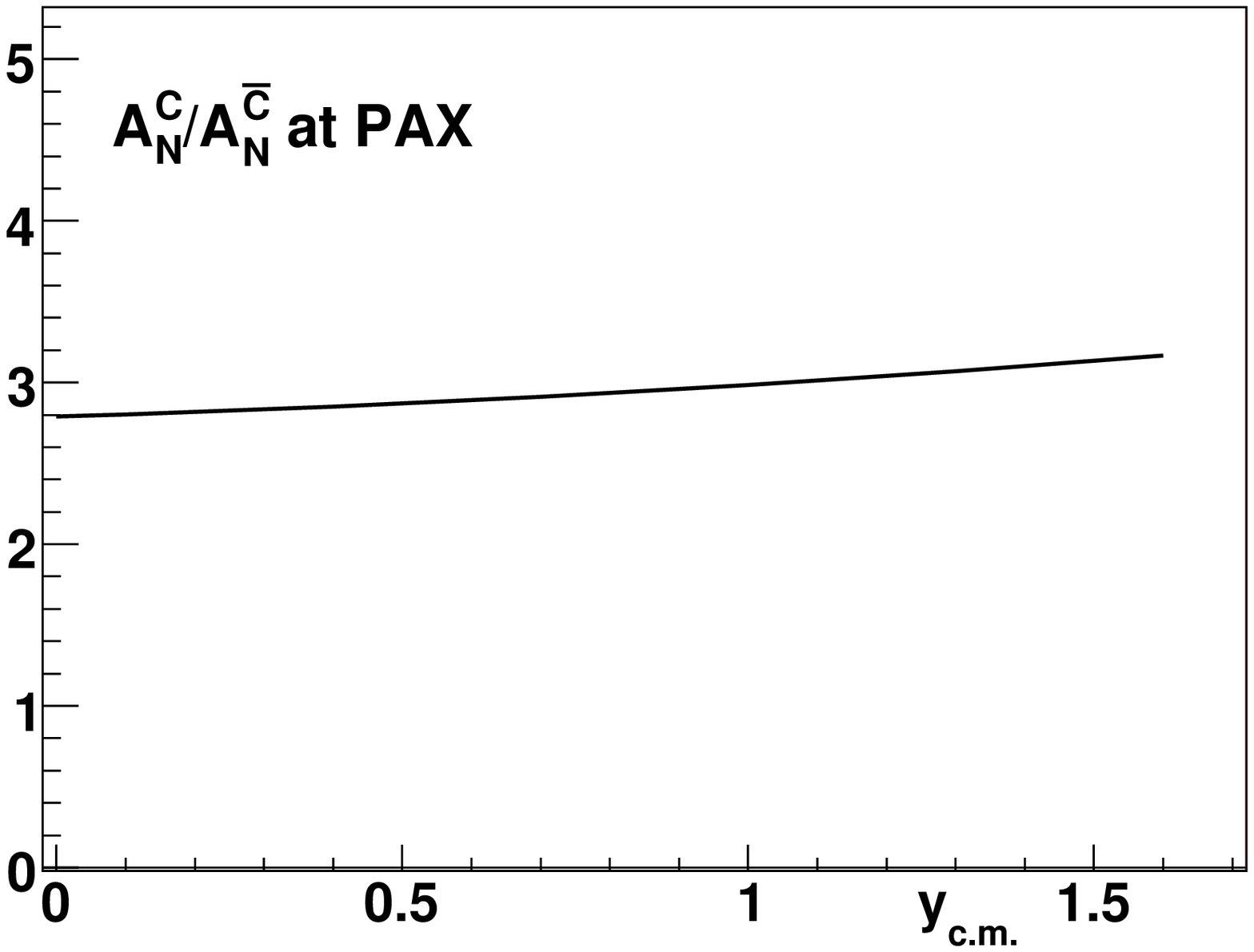}
\end{center}
\vskip -0.4cm \caption{\it Ratio of the SSA for the charm quark over
that for the anticharm quark as function of charm quark rapidity
in $p^\uparrow p$ scattering at JPARC (left) and
$\bar p^\uparrow p$ scattering at PAX (right) experiments.}
\end{figure}

The sizes of the above single spin asymmetries also depend on
the dominance of the quark-antiquark annihilation channel at both energies.
In our leading order simulation with the CTEQ5L parameterization for the
parton distributions, this channel indeed dominates the cross sections.
These asymmetries may scale down if the gluon-gluon channel turns dominant~\cite{nlo}.
However, we shall still observe the big difference between the SSAs for
charm and anticharm quarks when the twist-three gluon-gluon correlation contribution
to the SSAs~\cite{Ji:1992eu,newjw} is small in the valence region.
In order to demonstrate this difference, we plot in Fig.~4 the ratios of the
asymmetries as functions of the rapidities at both experiments. From
these plots, we find that the ratio at both experiments is about
a factor of 3 and has little dependence on the rapidity. These
ratios are model-independent predictions, especially at forward
region of the polarized nucleon where the quark-antiquark channel dominates.
We hope the experiments at both facilities can carry out these measurements,
and test the predictions.
We further notice that the sea quark contribution in the quark-gluon correlation functions
from the polarized nucleon will be
important at the central rapidities~\cite{new}, where the above ratios may vary if
we take into account those contributions.

{\bf 4. Summary.} In this paper, we studied the single spin
asymmetries in heavy quark and antiquark production in hadronic
processes. Because of the different color charges, the final state
interactions on the heavy quark and antiquark are different, and
therefore the associated single spin asymmetries are different too.
Our numerical simulations showed that the single spin asymmetry
for heavy quark is about a factor of 3 larger than that for the antiquark
in the forward $p^\uparrow p$ amd $p^\uparrow \bar p$ scattering at
low energy sacttering where the quark-antiquark channel dominates.
The experimental observation of this unique signature shall provide
a crucial test for the underlying physics for the single spin asymmetry
phenomena.

In the above analysis, we only considered the quark-antiquark channel
contributions. An extension to the gluon-channel contribution can
follow accordingly, which will be relevant for the collider experiment at
RHIC~\cite{Anselmino:2004nk,{Ji:1992eu},newjw}.
We reserve the study of this contribution as well as other contribution, for example,
from the chiral-odd quark-gluon correlation functions~\cite{Kanazawa:2000hz}
in a future publication. We further notice that the latter contribution will have
the similar pattern for the SSAs in heavy quark and antiquark production
as the chiral-even one as we discussed in this paper, because of the final
state interaction effects dominate both contributions.

This work was supported in part by the U.S. Department of Energy under contract
DE-AC02-05CH11231. We are grateful to RIKEN, Brookhaven National
Laboratory and the U.S. Department of Energy (contract number
DE-AC02-98CH10886) for providing the facilities essential for the
completion of this work. J.Z. is partially supported by China Scholarship Council and 
National Natural Science Foundation of China under Project No. 10525523.

\end{document}